# Hydrogen Geysers: Explanation for Observed Evidence of Geologically Recent Volatile-Related Activity on Mercury's Surface

J. Marvin Herndon
Transdyne Corporation
11044 Red Rock Drive
San Diego, CA 92131 USA
mherndon@san.rr.com

Many of the images from the MESSENGER spacecraft reveal "… an unusual landform on Mercury, characterized by irregular shaped, shallow, rimless depressions, commonly in clusters and in association with high-reflectance material … and suggest [1] that it indicates recent volatile-related activity" (Figures 1 and 2). But the authors were unable to describe a scientific basis for the source of those volatiles or to suggest identification of the "high-reflectance material".

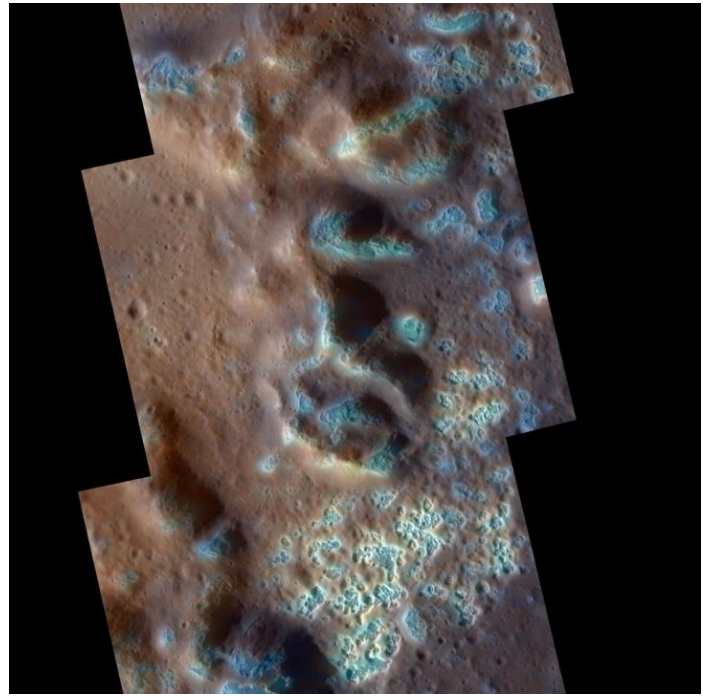

**Figure 1.** Colorized NASA MESSENGER image, taken with the Narrow Angle Camera, shows an area of hollows (blue) on the floor of Raditladi basin on Mercury. Surface hollows were first discovered on Mercury during MESSENGER's orbital mission and have not been seen on the Moon or on any other rocky planetary bodies. These bright, shallow depressions appear to have been formed by disgorged volatile material(s) from within the planet.

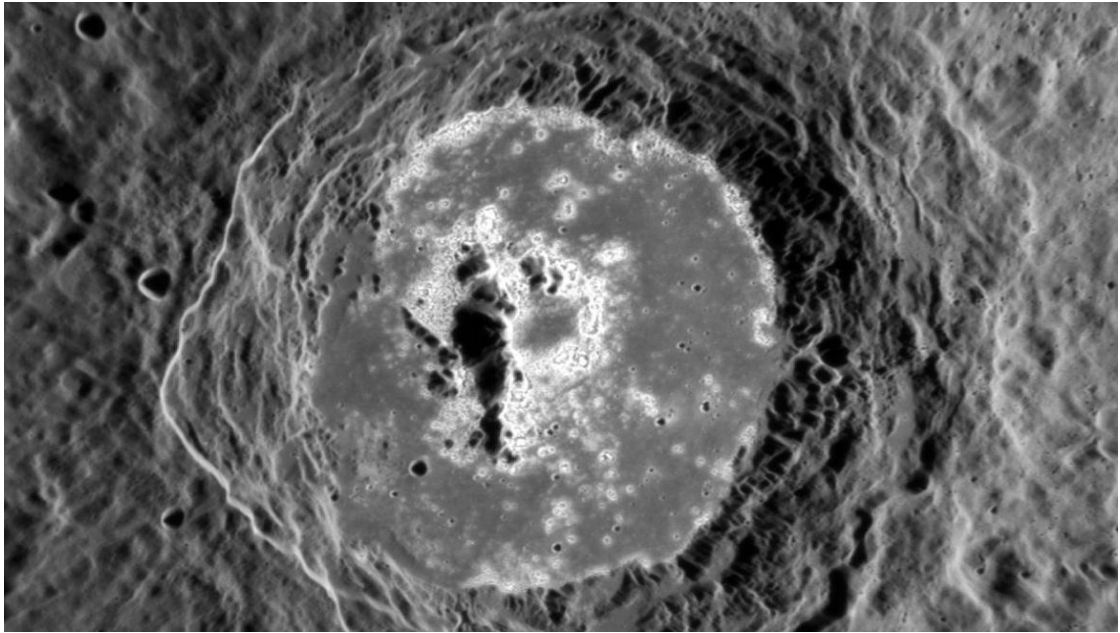

**Figure 2.** NASA MESSENGER image of a complex crater exhibits many hollows along its floor and central peak complex. The hollows have a very high albedo, which makes this crater stand out prominently.

I agree with the authors' interpretation of volatile activity being responsible forming those "unusual landforms" and agree that "…Mercury's interior contains higher abundances of volatile elements than are predicted by several planetary formation models for the innermost planet." Here I describe a basis of planetary formation, not considered by the authors and unlike their cited models, and show that primordial condensation from an atmosphere of solar composition at pressures of one atmosphere or above leads to the incorporation of copious amounts of hydrogen in Mercury's core, much of which is released as the core solidifies. The release of hydrogen escaping at the surface, I posit, is responsible for the formation of said "unusual landform on Mercury", sometimes referred to as pits, and for the formation of the associated "high-reflectance material", bright spots, which I suggest is iron metal reduced from an iron compound, probably iron sulfide, by the escaping hydrogen.

Thermodynamic considerations led Eucken [2] to conceive of Earth formation from within a giant, gaseous protoplanet where molten-iron rained out to form the core, followed by the condensation of the silicate-rock mantle. By similar, extended calculations I verified Eucken's results and deduced that oxygen-starved, highly-reduced matter characteristic of enstatite chondrites and, by inference, also the Earth's interior (Table 1), condensed at high temperatures and high pressures from primordial Solar System gas under circumstances that isolated the condensate from further reaction with the gas at low temperatures [3, 4], circumstances that I extend here to planet Mercury.

Ideally, in a cooling atmosphere of solar composition, iron starts to condense when the partial pressure of iron gas exceeds the vapor pressure of iron metal [5], $P^V$ (Fe), according to

$$[ 2A_{Fe} / A_H ] P(H_2) = P^V (Fe) \tag{1}$$

Where the A's are primordial elemental abundance ratios [6] and the pressure of hydrogen gas, $H_2$, is approximately equal to the total pressure. Thus, at higher $H_2$-pressures, iron can condense at higher temperatures, even the temperatures at which iron is liquid. Hydrogen is readily soluble in molten iron, where for an ideal solution, the solubility of hydrogen, $C_H$, in mL per 100 g. of iron is given by [7]

$$\ln C_H = 5.482 - 4009/T + \tfrac{1}{2} \ln [ P(H_2)/P(\text{reference} = 1 \text{ atm.}) ] \tag{2}$$

The solid curve in Figure 3 shows the range of temperatures and pressures at which molten iron will ideally begin to condense from an atmosphere of primordial (solar) composition, calculated from equation (1). The dashed curve in Figure 3 shows the amount of hydrogen that could ideally dissolve in Mercury's molten iron (estimated at one-half Mercury's mass) in equilibrium with primordial hydrogen gas, calculated at points along the solid curve and expressed as volume of dissolved $H_2$ at STP [standard temperature and pressure, 293K, 1 atm.] relative to the volume of planet Mercury. These calculations are not intended to model Mercury's formation; too many unknowns are involved for precise determinations, such which alloying elements enhance or de-enhance gas solubility, or which precise range of temperatures might be involved. Rather, the calculations are intended to illustrate within a broad range of conditions that planetary condensation at $H_2$-pressures of one atmosphere or above can lead to copious amounts of hydrogen incorporated in Mercury's core, which as it subsequently solidifies, will be released.

The release of dissolved hydrogen during Mercury's core solidification is, by Figure 3, certainly sufficient in amount to account for the "unusual landform" on Mercury's surface and is possibly involved in the exhalation of iron sulfide, which is abundant on the planet's surface, and some of which may have been reduced to iron metal thus accounting for the associated "high-reflectance material", bright spots. So, here is a test. Proving that the "high-reflectance material" is indeed metallic iron will provide strong evidence that the exhausted gas is hydrogen and evidence of the basis of planetary formation at pressures at or above about one atmosphere as described here; a negative result, however, would not disprove hydrogen disgorgement and might suggest the highly reflective material is enstatite, $MgSiO_3$.

I have suggested that only three processes, operant during the formation of the Solar System, are primarily responsible for the diversity of matter in the Solar System and are directly responsible for planetary internal compositions and structures [3]. These are: (*i*) High-pressure, high-temperature condensation from primordial matter associated with planetary-formation by raining out from the interiors of giant-gaseous protoplanets; (*ii*) Low-pressure, low-temperature condensation from primordial matter in the remote reaches of the Solar System or in the interstellar medium; and, (*iii*) Stripping of the primordial volatile components from the inner portion of the Solar System by super-intense T-Tauri phase outbursts, presumably during the thermonuclear ignition of the Sun. Verifying that the "high-reflectance material" is indeed metallic iron will not only provide strong evidence for Mercury's hydrogen geysers, but more generally will provide evidence that planetary interiors "rained out" by condensing at high pressures within giant-gaseous protoplanets.

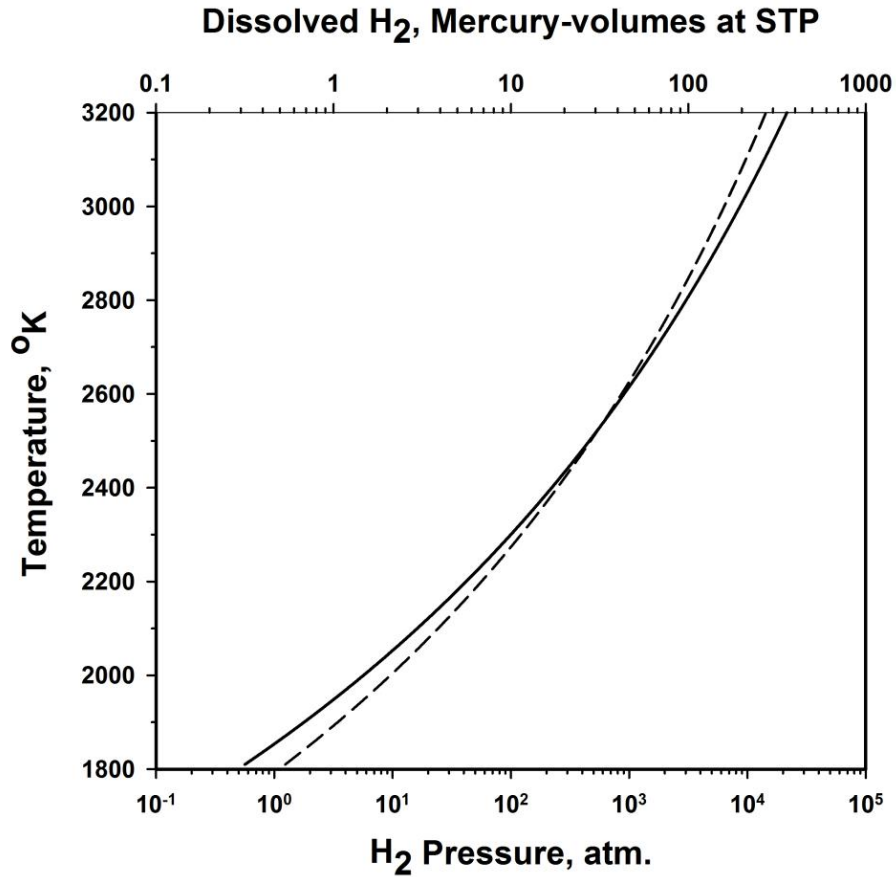

**Figure 3.** The solid curve in this figure shows the temperatures and hydrogen pressures in a cooling atmosphere of solar composition at which liquid iron will begin to condense. The dashed curve the substantial volume of hydrogen ideally that could dissolve in Mercury's iron in that medium at the temperatures and pressures indicated by the solid curve. Dissolved hydrogen is expressed as volume at standard temperature and pressure, and expressed as multiples of planet Mercury's volume.

**Table 1.** Fundamental mass ratio comparison between the endo-Earth (lower mantle plus core) and the Abee enstatite chondrite. Above a depth of 660 km seismic data indicate layers suggestive of veneer, possibly formed by the late addition of more oxidized chondrite and cometary matter, whose compositions cannot be specified at this time.

| Fundamental Earth Ratio | Earth Ratio Value | Abee Ratio Value |
|---|---|---|
| lower mantle mass to total core mass | 1.49 | 1.43 |
| inner core mass to total core mass | 0.052 | theoretical 0.052 if $Ni_3Si$ 0.057 if $Ni_2Si$ |
| inner core mass to lower mantle + total core mass | 0.021 | 0.021 |
| D″ mass to total core mass | 0.09‡ | 0.11* |
| ULVZ† of D″ CaS mass to total core mass | 0.012⌐ | 0.012* |

\* = avg. of Abee, Indarch, and Adhi-Kot enstatite chondrites
D″ is the "seismically rough" region between the fluid core and lower mantle
† ULVZ is the "Ultra Low Velocity Zone" of D″
‡ calculated assuming average thickness of 200 km
⌐ calculated assuming average thickness of 28 km
data from [8-10]

## Acknowledgements

I thank Robert D. Pehlke (University of Michigan), and James MacAllister and the late Lynn Margulis (University of Massachusetts, Amherst) for beneficial discussions.